\documentstyle[12pt]{article}
\renewcommand{\theequation}{\arabic{section}.\arabic{equation}}
\begin{document}
\title{
        Strings and $p$-branes with or without spin
        degrees of freedom and $q$-form fields
        }

\author{
        Tadashi MIYAZAKI  \\
        {\it Department of Physics, Science University of Tokyo,} \\
        {\it Kagurazaka, Shinjuku-ku, Tokyo 162, Japan}
}

\date{}

\maketitle
\vspace{20mm}

\begin{abstract}
A general action is proposed for the fields of $q$-dimensional differential form
over the compact Riemannian manifold of arbitrary dimensions.
Mathematical tools come from the well-known 
de Rham-Kodaira decomposing theorem on the harmonic integral. 
We have a field-theoretic action suitable for strings and $p$-branes with or 
without spin degrees of freedom.
In a completely-kinematical way is derived the generalized Maxwell theory with
a magnetic monopole over a curved space-time, where we have a new 
type of gauge transformations.

\end{abstract}

\setlength{\baselineskip}{7.8mm}
\newpage

\section{Introduction}
\hspace*{6mm}
It goes without saying that field theories play a central role in drawing a 
particle picture. They are especially important to explore a way to construct
a theoretical view on a $curved$ space-time (of more than four dimensions).
Recently-developed theories of strings\cite{1} and membranes\cite{2}, as well 
as those
of two-dimensional gravity\cite{3}, go along this way. 
If one completes a picture with a general action, 
one may have a clear understanding about why the fundamental structure is of 
one dimension (a string), excluding other extended structures
of two or more dimensions.

The first purpose of this paper is to obtain a general action for
the fields of $q$-dimensional differential forms on a general
curved space-time.
In such a way can we deal not only with strings and $p$-branes
( $p$-dimensional extended objects), 
but also with vector and tensor fields as assigned on each point of a
compact Riemannian manifold (e.g., a sphere or a torus of
general dimensions).
 
Our next aim is, as a result of this treatment, to generalize the
conventional Maxwell theory to that on the $curved$ space-time
of {\it arbitrary dimensions} .
Our method is based on the mathematical theory having been developed by de 
Rham and Kodaira\cite{4}. In the theory of harmonic integrals
the elegant theorem,
having been now crowned with the names of the two brilliant mathematicians,
says that an arbitrary differential form consists of three parts:
a harmonic form, a $d$-boundary and a $\delta$-boundary. 
With this theorem we derive a general Maxwell theory only $kinematically$ , i.
e.,
through mathematical manipulation. 
We have an electromagnetic field coming from the $d$-boundary,
whereas a magnetic
monopole field from the $\delta$-boundary. We are thus to have a generalized
Maxwell theory with an electric charge and a magnetic monopole on an 
arbitrary-dimensional curved space-time.

In this paper we proceed by taking various concrete examples to construct a
field theory. Section 2 treats an algebraic method for obtaining a general act
ion.
Sections 3 to 5 are devoted to concrete examples. In Sect.6 we put concluding 
remarks and summary. Two often-used mathematical formulas are listed in 
Appendix.

We hope the method developed here will become one of the steps which one makes
forward to construct the field theory of all extended objects \,\,\,-----\,\,\, strings and $p$\,-branes with or without 
{\it spin degrees of freedom} \,\,\,-----\,\,\, based on algebraic geometry.
\vspace{10mm}

\setcounter{equation}{0}
\setcounter{section}{1}
\section{A general action with $q$-forms}
\hspace*{6mm}
Let us start with a Riemannian manifold $M^n$, where we, the observers, live,
and with a submanifold $\bar{M}^m$, where particles live ( $n,m$ : dimension o
f
the spaces; $n\geq m$). Both $M^n$ and $\bar{M}^m$ are
supposed to be $compact$
\,\,\,-----\,\,\, compact only because mathematicians construct
a beautiful theory of
harmonic forms over $compact$ spaces, and de Rham-Kodaira's theorem or Hodge's
theorem has not yet been proven with respect to the differential forms over
{\it non-compact} spaces.

We will admit the space $\bar{M}^m$ of a particle to be a submanifold of $M^n$. For
instance, $\bar{M}^m$ may be a circle or a sphere within an $n$-dimensional
(compact) space $M^n$. The local coordinate systems of $M^n$ and $\bar{M}^m$
shall be denoted by $(x^{\mu})$ and $(u^i)$, respectively 
[$\mu =1,2,...,n; i=1,2,...,m$]\cite{5}.  A point ( $u^1$,$u^2$,...,$u^m$) of 
$\bar{M}^m$ is, at the same time, a point of $M^n$,
so that it is also expressed
by $x^\mu=x^\mu(u^i)$. 
In a conventional quantum field theory, point particles, scalar fields, vector
or higher-rank tensor fields, or spinor fields are attributed to each point
of $\bar{M}^m$. In this view we are to assign a $q$-dimensional differential
form ({\it q-form}) $F^{(q)}$ to each point of $\bar{M}^m$, which is expressed
, as
mentioned above, by the local coordinate $(u^1,u^2,...,u^m)$ or by 
$x^{\mu}=x^{\mu}(u^i)$. 
Physical objects \,\,\,-----\,\,\, point particles, strings or
electromagnetic fields \,\,\,-----\,\,\,
should be identified with these $q$-forms.

We then make an action with $F^{(q)}$. One of the candidates
for the action $S$
is $(F^{(q)} ,F^{(q)}) \equiv \int_{\bar{M}^m}F^{(q)}*F^{(q)}$, where $*$
means Hodge's star operator transforming a $q$-form into 
an $(m-q)$-form. Expressed with respect to
an {\it orthonormal basis} $\omega _1,\omega _2,...,\omega _m$, it becomes 
\begin{equation}
*(\omega _{i_1} \wedge \omega _{i_2} \wedge ... \wedge \omega _{i_q})
= \frac{1}{(m-q)!} \delta\left( \begin{array}{cccccc}
1 & 2 & ... & ... & ... & m \\ i_1 & ... &  i_q , &  j_1 & ... & j_{m-q}
\end{array}
\right) 
\omega _{j_1} \wedge \omega _{j_2} \wedge ... \wedge \omega _{j_{m-q}}, 
\label{2:1}
\end{equation}
where $\delta (....)$ denotes the signature $(\pm)$ of the permutation and
the summation convention over repeated indices is , here and hereafter, always
implied. The inner product $(F^{(q)},F^{(q)})$ is a scalar and shares a 
property of scalarity with the action $S$.
Let us , therefore, admit the action $S$ to be proportional to 
$(F^{(q)},F^{(q)})$ and investigate each case that we confront with in the
conventional theoretical physics. 
Thus we put
\begin{eqnarray}
S &=& (F^{(q)},F^{(q)}) = \int_{\bar{M}^m}\hspace{-2mm}{\cal S} \nonumber \\
&=& \int_{\bar{M}^m}\hspace{-2mm}{\cal L}\hspace{1mm}
du^1\wedge du^2\wedge ...\wedge du^m,
\label{2:2} \\
{\cal S} &\equiv& F^{(q)}\!*F^{(q)} = {\cal L}\hspace{1mm}
du^1 \wedge du^2 \wedge  ... \wedge du^m.
\nonumber
\end{eqnarray}
Here ${\cal S}$ is an {\it action form}, but we will sometimes call it by the 
same name $action$. ${\cal L}$ is interpreted as a Lagrangian density.

According to the well-known de Rham-Kodaira theorem, an arbitrary $q$-form
decomposes into the three mutually orthogonal $q$ -forms:
\begin{eqnarray}
F^{(q)}=F_{\rm I}^{(q)} + F_{\rm II}^{(q)} + F_{\rm III}^{(q)} ,
\label{2:3}
\end{eqnarray}
where $F_{\rm I}^{(q)}$ is a harmonic form, meaning\cite{6}
\begin{eqnarray}
dF_{\rm I}^{(q)} = \delta F_{\rm I}^{(q)} = 0,
\label{2:4}
\end{eqnarray}
and $F_{\rm II}^{(q)}$ is a $d$-boundary, and $F_{\rm III}^{(q)}$ is a 
$\delta$-boundary (coboundary). Here $\delta$ is Hodge's adjoint operator,
which implies $\delta = (-1)^{m(q-1)+1}*d*$ when operated to $q$-forms over 
the $m$-dimensional space. There exist, therefore, a $(q-1)$-form 
$A_{\rm II}^{(q-1)}$ and a $(q+1)$-form $A_{\rm III}^{(q+1)}$, such that
\begin{eqnarray}
F_{\rm II}^{(q)} = dA_{\rm II}^{(q-1)} \: ; \: F_{\rm III}^{(q)} = 
\delta A_{\rm III}^{(q+1)}.
\label{2:5}
\end{eqnarray}
The action $S$ is (proportional to) $(F^{(q)},F^{(q)})$ ;
\begin{eqnarray}
{\cal S} &\equiv& (F^{(q)},F^{(q)}) \nonumber \\
&=& (F_{\rm I}^{(q)},F_{\rm I}^{(q)}) + (A_{\rm II}^{(q-1)},
\delta dA_{\rm II}^{(q-1)}) + (A_{\rm III}^{(q+1)},d\delta A_{\rm III}^{(q+1)}),
\label{2:6} \\
S &\equiv& \int _{\bar{M}^m}\hspace{-2mm}{\cal S} = \int\!{\cal L}
\hspace{1mm}du^1 \wedge du^2 \wedge .... \wedge du^m. \nonumber
\end{eqnarray}
The physical meaning of Eq.(\ref {2:6}) is whatever we want to discuss in
this paper and
will be described in detail from now on.
\vspace{10mm}

\setcounter{equation}{0}
\section{Example 1.\,\,point particles, strings and $p$-branes}

\hspace*{6mm}
We first assign $F^{(0)}=1$ to a point $(u^1,...,u^m)$ of the submanifold 
$\bar{M}^m$, and we always make use of the relative (induced) metric 
$\bar{g}_{ij}$ for $\bar{M}^m$ (so that the intrinsic metric of the 
submanifold is irrelevant).
\begin{eqnarray}
\bar{g}_{ij} \equiv \frac{\partial x^\mu(u)}{\partial u^i}
\frac{\partial x^\nu(u)}{\partial u^j} g_{\mu\nu},
\label{3:1}
\end{eqnarray}
where  $g_{\mu\nu}$ is a metric of the Riemannian space $M^n$.
Since the volume element
$dV \equiv \omega_1\wedge \omega_2\wedge...\wedge \omega_m$ is 
expressed, with respect to the local coordinate $(u^i)$, as
\begin{eqnarray}
dV = \sqrt{\bar{g}}\hspace{1mm}du^1\wedge du^2\wedge...\wedge du^m = *1,
\label{3:2}
\end{eqnarray}
we immediately find
\begin{eqnarray}
(F^{(0)},F^{(0)})= \int_{\bar{M}^m}\hspace{-2mm}\sqrt{\bar{g}}
\hspace{1mm}du^1\wedge du^2\wedge...\wedge du^m,
\label{3:3}
\end{eqnarray}
with $\bar{g}=\det (g_{ij})$.

When $n=4$ and $m=1$, we have
\begin{eqnarray}
\bar{g} = g_{\mu\nu} \frac{dx^\mu}{du^1} \frac{dx^\nu}{du^1} = g_{\mu\nu} \dot
{x}^\mu \dot{x}^\nu, 
\label{3:4}
\end{eqnarray}
( $\cdot$ means $d/du^1$), hence
\begin{eqnarray}
(F,F) &=&\int_{\bar{M}^1}\hspace{-2mm}ds , 
\label{3:5} \\
ds^2 &=&  g_{\mu\nu} \dot{x}^\mu \dot{x}^\nu {(du^1)}^2 = g_{\mu\nu} dx^{\mu} 
dx^{\nu} , \nonumber
\end{eqnarray}
which indicates that $(F,F)$ is an action (up to a constant) for a point particle in
a curved 4-dimensional space, with $u^1$, interpreted as a proper time.

On the contrary, if we take up a submanifold $\bar{M}^2$, Eq.(\ref{3:3})
becomes
\begin{eqnarray}
(F^{(0)}, F^{(0)}) = \int _{\bar{M}^2}\hspace{-2mm}\sqrt{\bar{g}}
\hspace{1mm}du^1 \wedge du^2\,,
\label{3:6}
\end{eqnarray}
with
\begin{eqnarray}
\bar{g} = \det ( \frac{\partial x^{\mu}}{\partial u^i} \frac{\partial x^{\nu}}
{\partial u^j} g_{\mu\nu}),
\label{3:7}
\end{eqnarray}
which is just the Nambu-Goto action in a $curved$ space (with $u^1=\tau$ and $
u^2=\sigma$ in a conventional notation).
There, and here, the determinant $\bar{g}$ of an induced metric plays an
essential role.
If we confront with an arbitrary submanifold $\bar{M}^{p+1}$
($p$ : an arbitrary integer $\leq n-1$,
we are to have a $p$-brane, whose action is nothing but
that given by Eq.(\ref{3:3}) with $m=p+1$.

Let us discuss the transformation property of the action or Lagrangian density
.
The transformation of $\bar{M}^m$ into $\bar{M'}^m$ without changing 
$M^n$\cite{7} means reparametrization.
\begin{eqnarray}
u^i &\rightarrow& u'^i ,
\label{3:8} \\
x^{\mu}(u^i) &\rightarrow& x'^{\mu}(u'^i) = x^{\mu}(u^i). \nonumber
\end{eqnarray}
By this the volume element Eq.(\ref{3:2}) does not change, so that our
Lagrangian (density) for the $p$ -brane is trivially invariant under the 
reparametrization.
If we convert $M^n$ into $M'^n$ without changing $\bar{M}^m$, a general
coordinate transformation
\begin{eqnarray}
x^{\mu}(u^i) \rightarrow x'^{\mu}(u^i)  
\label{3:9}
\end{eqnarray}
is induced, under which $\bar{g}_{ij}$ does not change, because of the 
transformation property of the metric $g_{\mu\nu}$.
Our action is trivially invariant also for this general coordinate 
transformation.

If we transform $\bar{M}^m$ and $M^n$ simultaneously, i.e.,
\begin{eqnarray}
u^i &\rightarrow& u'^i , \nonumber \\
x^{\mu}(u^i) &\rightarrow& x'^{\mu}(u'^i),
\label{3:10}
\end{eqnarray}
we do not have an equality $x'^{\mu}(u'^i) = x^{\mu}(u^i)$.
This type of transformations is examined, as an example, for $n=3$ and $m=2$
as follows. Let us take $M^3={\sf R}^3$ (compactified), and
$\bar{M}^2={\sf S}^2$ (2-dim surface of a sphere)
whose local coordinate system
is $(u^1 , u^2)$. A point of ${\sf S}^2$ is expressed by $(u^1 , u^2)$, but
it is at the same time a point $(x^1 , x^2, x^3)$ of ${\sf R}^3$. 
We give the relation between the two coordinate systems by the
stereographic projection:
\begin{eqnarray}
x^1 &=& \frac{2r^2 u^1}{(u^1)^2 + (u^2)^2 + r^2}\,, \nonumber \\
x^2 &=& \frac{2r^2 u^2}{(u^1)^2 + (u^2)^2 + r^2}\,, \\
x^3 &=& \frac{r[r^2-(u^1)^2 - (u^2)^2]}{(u^1)^2 + (u^2)^2 + r^2}\,, \nonumber 
\label{3:11}
\end{eqnarray}
where $r$ is the radius of the sphere defining ${\sf S}^2$. The transformation
$(u^1, u^2) \rightarrow (u'^1,u'^2)$ induces the transformation 
$(x^1, x^2, x^3) \rightarrow (x'^1, x'^2, x'^3)$, and vice versa.
The definition of the metric $g_{\mu\nu}$ for $M^n$ and the induced one 
$\bar{g}_{ij}$ for $\bar{M}^m$ tells us 
\begin{eqnarray}
g'_{\mu\nu}(x') = \frac{\partial x^{\rho}}{\partial x'^{\mu}} \frac{\partial x
^{\delta}}{\partial x'^{\nu}} g_{\rho\delta} (x) , 
\label{3:12}
\end{eqnarray}
and
\begin{eqnarray}
\bar{g}'_{ij}(u') = \frac{\partial u^k}{\partial u'^i} \frac{\partial u^l}
{\partial u'^j} \bar{g}_{kl} (u) , 
\label{3:13}
\end{eqnarray}
so that we have
\begin{eqnarray}
\sqrt{\bar{g}'(u')}\hspace{1mm}du'^1 \wedge ... \wedge du'^m = \sqrt{\bar{g}(u
)} \hspace{1mm}du^1 \wedge ... \wedge du^m ,
\label{3:14}
\end{eqnarray}
hence follows the invariance of the action.
\vspace{10mm}

\setcounter{equation}{0}
\section{Example 2.\,\,scalar fields}

\hspace*{6mm}
Now we consider the case where a scalar field $\phi (x^{\mu}(u^i))$ is 
assigned to each point $x^{\mu}(u^i)$. From now on we regard every 
quantity as that given over the subspace $\bar{M}^m$, hence we will write
the field simply as $\phi (u^i)$ instead of $\phi (x^{\mu}(u^i))$ , etc.

An arbitrary 0-form \,\,\,-----\,\,\, a scalar field \,\,\,-----\,\,\,
decomposes into two parts: 
\begin{eqnarray}
F^{(0)} = F_{\rm I}^{(0)} + F_{\rm III}^{(0)}.
\label{4:1}
\end{eqnarray}
$F_{\rm I}^{(0)}$ is given by
\begin{eqnarray}
F_{\rm I}^{(0)} = \phi (u) ,
\label{4:2}
\end{eqnarray}
with which we obtain 
\begin{eqnarray}
(F_{\rm I}^{(0)} , F_{\rm I}^{(0)}) = \phi ^2(u) dV ,  
\label{4:3}
\end{eqnarray}
meaning a mass term of a scalar field. $F_{\rm III}^{(0)}$ is composed, on the
contrary , of a $\delta$ -boundary of a 1-form:
\begin{eqnarray}
F_{\rm III}^{(0)} &=& \delta A^{(1)} , \nonumber \\
A^{(1)} &=& A_i du^i .
\label{4:4}
\end{eqnarray}
Hence we have
\begin{eqnarray}
F_{\rm III}^{(0)} = -\partial _k (\sqrt{\bar{g}} A^k) \sqrt{\bar{g}} \bar{g} ^
{11} \bar{g}^{22} ... \bar{g}^{mm} ,
\label{4:5}
\end{eqnarray}
where, as usual,
\begin{eqnarray}
A^k = \bar{g}^{kl} A_l \:\: {\rm and} \:\: \partial _k = \frac{\partial}
{\partial u^k} ,
\label{4:6}
\end{eqnarray}
and $(\bar{g} ^{ij})$ is the inverse of $(\bar{g} _{ij})$. In a special
case, where we work with a flat space and an orthonormal basis, i.e.,
\begin{eqnarray}
\bar{g} ^{ij} = \delta ^{ij} \:\: {\rm and} \:\: du^i = \omega ^i , 
\label{4:7}
\end{eqnarray}
we have a simple form
\begin{eqnarray}
F_{\rm III} ^{(0)} = - \partial _k A^k  ,  
\label{4:8}
\end{eqnarray}
by which the action form ${\cal S}$ becomes
\begin{eqnarray}
{\cal S} = (\partial _k A^k)^2 dV . 
\label{4:9}
\end{eqnarray}
This is the `kinetic' term of the {\it k-vector field} $A^k$.

The {\it gauge transformation} exists for this field: 
\begin{eqnarray}
A^{(1)} &\rightarrow& \tilde{A}^{(1)} = A^{(1)}+\delta A^{(2)} , \nonumber \\
A^{(2)} &=& \frac{1}{2} A_ {{i_1}{i_2}}du^{i_1}\wedge du^{{i_2}} .
\label{4:10}
\end{eqnarray}
In components, it is written as
\begin{eqnarray}
\tilde{A} _h\! =\! A_h\!\! +\! \frac{1}{2(m-2)!} \delta\!\! \left( \begin{array}{ccccc}
h & l_1 & ... & ... & l_{m-1} \\
i_1 & i_2 & j_1 & ... & j_{m-2} 
\end{array}
\right )\!
\frac{\partial (\sqrt{\bar{g}} A^{i_1 i_2})}{\partial u^k} 
\sqrt{\bar{g}}\bar{g}^{k l_1}\bar{g}^{l_2 j_1}\!...
\bar{g}^{l_{m-1} j_{m-2}} . 
\label{4:11}
\end{eqnarray}
One can further calculate , if one wants to , to have a beautiful form:
\begin{eqnarray}
\tilde{A} _i &=& A_i - \frac{1}{2} \delta \left( \begin{array}{cc}
j_1 & j_2 \\
k & i 
\end{array}
\right )
\bar{g} ^{kl} D_l A_{j_1 j_2}, \nonumber \\
D_l A_{j_1 j_2} &=& \frac{\partial A_{{j_1}{j_2}}}{\partial u^l} - A_{k j_2} \
Gamma_{{j_1}l}^k-A_{{j_1}k} \Gamma_{{j_2}l}^k , 
\label{4:12}
\end{eqnarray}
where $\Gamma_{jk}^i$ is the well-known affine connection.
\begin{eqnarray}
\Gamma_{jk}^i = \frac{1}{2} g^{il} (\frac{\partial \bar{g} _{jl}}{\partial u^k
} + \frac{\partial \bar{g} _{lk}}{\partial u^j} - \frac{\partial \bar{g} _{jk}
}{\partial u^l} ) .
\label{4:13}
\end{eqnarray}
Note that our fundamental fields are the $A_i$, and the gauge transformation 
is obtained with the $A_{i_1 i_2}$ of the rank {\it higher by one} than the 
former.
This is, of course, due to the nilpotency of $\delta$, $\delta ^2 = 0$,
and typical
of our new type of formulation.
\vspace{10mm}

\setcounter{equation}{0}
\section{Example 3.\,\,vector fields}

\hspace*{6mm}
When a 1-form $F^{(1)}$ is assigned to each point of $\bar{M}^m$, we have
\begin{eqnarray}
F^{(1)} = F_{\rm I}^{(1)} + F_{\rm II}^{(1)} + F_{\rm III}^{(1)} .
\label{5:1}
\end{eqnarray}
First we will see the contribution of $F_{\rm I}^{(1)}$ to the action,
which is harmonic. Writing as 
\begin{eqnarray}
F_{\rm I}^{(1)} = F_i du^i ,
\label{5:2}
\end{eqnarray}
we immediately have an action (form)
\begin{eqnarray}
{\cal S}_{\rm I} = F_{\rm I}^{(1)}*F_{\rm I}^{(1)}=F_iF^i\!
\sqrt{\bar{g}}\hspace{1mm}
du^1 \wedge ... \wedge du^m 
\label{5:3}
\end{eqnarray}

The contribution of the $d$-boundary is calculated in the same way. Putting
\begin{eqnarray}
F_{\rm II}^{(1)} = dA^{(0)} , 
\label{5:4}
\end{eqnarray}
we have the action
\begin{eqnarray}
{\cal S}_{\rm II} = g^{ij} \partial _i A^{(0)} \partial _j A^{(0)}\!\sqrt{\bar
{g}}
\hspace{1mm}du^1 \wedge ... \wedge du^m , 
\label{5:5}
\end{eqnarray}
which expresses a massless scalar particle $A^{(0)}$. Freedom of the choice of
gauges does not here appear. 

The contribution of the $\delta$-boundary is, on the contrary, rather
complicated in calculation. If we put
\begin{eqnarray}
F_{\rm III} ^{(1)} &=& \delta A^{(2)} , \nonumber \\
A^{(2)} &=& \frac{1}{2} A_{i_1 i_2} du^{i_1} \wedge du^{i_2} ,
\label{5:6} \\
F_{\rm III} ^{(1)} &=& F_i du^i , \nonumber 
\end{eqnarray}
we have
\begin{eqnarray}
F_h &=& \delta \left( \begin{array}{ccccc}
h & l_1 & l_2 & ... & l_{m-1} \\
i_1 & i_2 & j_1 & ... & j_{m-2}
\end{array}
\right) 
\frac {\partial}{\partial u^k} ( \sqrt{\bar{g}}A^{i_1 i_2})\sqrt{\bar{g}} 
\,\bar{g}^{kl_1} \bar{g}^{j_1 l_2} ... \bar{g}^{j_{m-2} l_{m-1}} \nonumber \\
&=& - \frac{1}{2} \delta \left( \begin{array}{cc} j_1 & j_2 \\ k & h 
\end{array} \right)
\bar{g}^{kl} D_l A_{{j_1}{j_2}} , 
\label{5:7}
\end{eqnarray}
with $D_l$, defined in Eq.(\ref{4:13})\cite{8}. 
The action is
\begin{eqnarray}
{\cal S}_{\rm III} = F_i F^i\!\sqrt{\bar{g}}\hspace{1mm}
du^1 \wedge ... \wedge du^m . 
\label{5:8}
\end{eqnarray}

The gauge transformation is given in this case by 
\begin{eqnarray}
A^{(2)} &\rightarrow& \tilde{A}^{(2)} = A^{(2)} + \delta A^{(3)} ,
\nonumber \\
A^{(3)} &=& \frac{1}{3!} A_{i_1 i_2 i_3}du^{i_1} \wedge du^{i_2}
\wedge du^{i_3} ,
\label{5:9}
\end{eqnarray}
which trivially leads to the relation
\begin{eqnarray} 
F_3 ^{(1)} = \delta A^{(2)} = \delta \tilde{A}^{(2)} . 
\label{5:10}
\end{eqnarray}
When expressed in components, it is written as
\begin{eqnarray}
\tilde{A} _{h_1 h_2} = 
A_{h_1 h_2} \!\!\!&-&\!\!\! \frac{1}{3!(m-3)!} \delta
\left(
\begin{array}{cccccc}
i_1 & i_2 & i_3 & j_1 & ... & j_{m-3} \\
h_1 & h_2 & l_1 & ... & ... & l_{m-2} 
\end{array}
\right)
\frac{\partial}{\partial u^k} ( \sqrt{\bar{g}} A^{i_1 i_2 i_3}) \nonumber \\
&\times&\!\!\!\sqrt{\bar{g}}\,\bar{g}^{kl_1}\bar{g}^{j_1 l_2}...\bar{g}^{j_{m-
3} l_{m-2}}, 
\label{5:11}
\end{eqnarray}
where, of course, the components with superscript are related to those with
subscript in a conventional manner, as has been described repeatedly.
\begin{eqnarray}
A^{i_1 i_2 i_3} = \bar{g}^{i_1 j_1} \bar{g}^{i_2 j_2} \bar{g}^{i_3 j_3} A_{j_1
 j_2 j_3} .
\label{5:12}
\end{eqnarray}
We finally express Eq.(\ref{5:11}) in an elegant form.
\begin{eqnarray}
\tilde{A} _{h_1 h_2} ^{(2)} &=& A _{h_1 h_2} ^{(2)} - \frac{1}{3!} 
\delta \left(
\begin{array}{ccc}
j_1 & j_2 & j_3 \\
k & h_1 & h_2
\end{array}
\right)
\bar{g}^{kl} D_l A_{j_1 j_2 j_3} , \nonumber \\
D_l A_{j_1 j_2 j_3} &=& 
\frac{\partial A_{j_1 j_2 j_3}}{\partial u^l} - A_{kj_2 j_3} \Gamma _{j_1 l}^k
- A_{j_1 k j_3} \Gamma _{j_2 l}^k - A_{j_1 j_2 k} \Gamma _{j_3 l} ^k. 
\label{5:13}
\end{eqnarray}

Especially when the space-time is flat and one takes an orthonormal reference
frame, one has 
\begin{eqnarray}
F_i = -\frac{1}{2} \delta \left( \begin{array}{cc}
k & i \\
i_1 & i_2
\end{array}
\right)
\frac{\partial A^{i_1 i_2}}{\partial u^k} ,
\label{5:14}
\end{eqnarray}
which further reduces to a familiar form {\it for} $m=4$: 
\begin{eqnarray}
F^i &=& \partial _k A^{ik} \nonumber \\
{\cal S} &=& \partial _k A^{ik} \partial _l A^{il} dV.
\label{5:15}
\end{eqnarray}
The gauge transformation becomes in this case
\begin{eqnarray}
\tilde{A} _{i_1 i_2} = A_{i_1 i_2} - \partial _k A_{i_1 i_2 k}  .
\label{5:16}
\end{eqnarray}

Needless to say, the total action comes from adding ${\cal S}_{\rm I}$,${\cal 
S}_{\rm II}$ and ${\cal S}_{\rm III}$. 
A new type of gauge transformations Eq.(\ref{5:13}) appears, due to the
coboundary property of $F_{\rm III} ^{(1)}$.
\vspace{10mm}

\setcounter{equation}{0}
\section{Example 4.\,\,tensor fields}

\hspace*{6mm}
Now we come to the case where a 2-form is assigned to each point of 
$\bar{M}^m$, the case of which is most useful and attractive for future
development. 

A 2-form decomposes, as usual, into the following three:
\begin{eqnarray}
F^{(2)} = F_{\rm I} ^{(2)} + F_{\rm II} ^{(2)} + F_{\rm III} ^{(2)}\,.
\label{6:1}
\end{eqnarray}
The harmonic form $F_{\rm I} ^{(2)}$ is written with the components
$A_{ij}$ as follows:
\begin{eqnarray}
F_{\rm I} ^{(2)} = \frac{1}{2} A_{i_1 i_2} du^{i_1} \wedge du^{i_2} ,
\label{6:2}
\end{eqnarray}
from which we have
\begin{eqnarray}
{\cal S}_I = F_{\rm I} ^{(2)} * F_{\rm I} ^{(2)} = 
\frac{1}{2} A_{i_1 i_2} A^{i_1 i_2}\!\sqrt{\bar{g}}\hspace{1mm}
du^1 \wedge ... \wedge du^m . 
\label{6:3}
\end{eqnarray}

The contribution of the $d$ -boundary is expressed with our fundamental 1-form
$A^{(1)}$. 
\begin{eqnarray}
F_{\rm II} ^{(2)} = dA ^{(1)} .
\label{6:4}
\end{eqnarray}
This further reduces, when written in components, 
\begin{eqnarray}
F_{\rm II} ^{(2)} &=& \frac{1}{2} F_{i_1 i_2} du^{i_1} \wedge du^{i_2} ,
\nonumber\\
A^{(1)} &=& A_i du^i ,
\label{6:5}
\end{eqnarray}
to a familiar relation
\begin{eqnarray}
F_{ij} = \partial _i A_j - \partial _j A_i , 
\label{6:6}
\end{eqnarray}
$(\partial _i = \partial / \partial u^i )$, which shows that $F_{ij}$
is a field-strength.
The gauge transformation here is given by
\begin{eqnarray}
A^{(1)} \rightarrow \tilde{A} ^{(1)} = A^{(1)} + dA^{(0)} . 
\label{6:7}
\end{eqnarray}
Namely, it is expressed in components as
\begin{eqnarray}
\tilde{A}_i = A_i + \partial _i A(u), 
\label{6:8}
\end{eqnarray}
with  $A(u)$, an arbitrary scalar function, which is a familiar form in the
conventional Maxwell electromagnetic theory. The invariance of the
contribution to $F_{\rm II} ^{(2)}$ owes self-evidently, to the
nilpotency $d^2=0$. 

If we further put
\begin{eqnarray}
\delta F_{\rm II} ^{(2)} = \delta dA^{(1)} = J ,
\label{6:9}
\end{eqnarray}
we have
\begin{equation}
- \frac{1}{2} \frac{1}{(m-2)!} \delta \left( \begin{array}{ccccc}
h & l_1 & l_2 & ... & l_{m-1} \\
i_1& i_2& j_1& ...& j_{m-2} \end{array} \right)
\frac{\partial}{\partial u^k} ( \sqrt{\bar{g}} F^{i_1 i_2}) \sqrt{\bar{g}} 
\hspace{1mm}\bar{g} ^{l_1 k} \bar{g} ^{l_2 j_1} ...
\bar{g} ^{l_{m-1} j_{m-2}} = J_h .
\label{6:10}
\end{equation}
After some lengthy calculations we finally have the following beautiful
form. 
\begin{eqnarray}
- \frac{1}{2} \delta \left( \begin{array}{cc} i_1 & i_2 \\ j & h \end{array} 
\right) \bar{g}^{jl} D_l F_{i_1 i_2} = J_h  .
\label{6:11}
\end{eqnarray}

The covariant derivative $D_l$ is given in Eq.(\ref{4:12}).
Equation (\ref{6:10})
or (\ref{6:11}) takes a simple form for the $flat$ $m$-dimensional space,
expressed 
in an {\it orthonormal basis}.
\begin{eqnarray}
F_{ij},^j = J_i
\label{6:12}
\end{eqnarray}
This is nothing but the Maxwell equation in an $m$-dimensional space,
with $J_i$, interpreted as an electromagnetic current density. One 
therefore finds that Eq.(\ref{6:9}) or (\ref{6:11}) is the generalized
Maxwell equation in the {\it curved m-dimensional space}.

From the viewpoint of {\it action-at-a-distance}\cite{9}, vector fields are
composed of matter fields. 
Namely, vector fields can be traced by looking at the matter fields.
Our standpoint is, on the contrary, such that our fundamental objects
are vectors and we can trace the matter field by regarding the vector
fields as such and calculating the left-hand side of Eq.(\ref{6:12}) with
Eq.(\ref{6:6}).
Our electromagnetic current of the matter $J_i(u)$ is determined by 
the vector fields $A_i(u)$. 

Now comes the contribution of the $\delta$-boundary :
\begin{eqnarray}
F_{\rm III} ^{(2)} = \delta A^{(3)},
\label{6:13}
\end{eqnarray}
where $A^{(3)}$ is a 3-form. Expressed, as usual, in components
\begin{eqnarray}
F_{\rm III} ^{(2)} &=& \frac{1}{2} F_{i_1 i_2} du^{i_1} \wedge du^{i_2} ,
\nonumber \\
A^{(3)} &=& \frac{1}{6} A_{i_1 i_2 i_3} du^{i_1} \wedge du^{i_2}
\wedge du^{i_3} ,
\label{6:14}
\end{eqnarray}
Eq.(\ref{6:13}) leads us to 
\begin{eqnarray}
F_{h_1 h_2} \hspace{-2mm}&=&\hspace{-2mm}
-\frac{1}{6(m-3)!} \delta\!\left( \begin{array}{cccccc}
h_1& h_2& l_1& ...&...& l_{m-2} \\ i_1& i_2& i_3& j_1& ...& j_{m-3}
\end{array} \right) 
\nonumber \\
&&\hspace{20mm}\times
\frac{ \partial}{\partial u^k}(\sqrt{\bar{g}} A^{i_1 i_2 i_3}) 
\sqrt{\bar{g}}\hspace{1mm}
\bar{g}^{l_1 k} \bar{g}^{l_2 j_1}\!...\bar{g}^{l_{m-2} j_{m-3}}.
\label{6:15}
\end{eqnarray}
Along the same line already mentioned repeatedly we further have
\begin{eqnarray}
F_{i_1 i_2} = - \frac{1}{6} \delta \left( \begin{array}{ccc} 
j_1 & j_2 & j_3 \\ k & i_1 & i_2 \end{array} \right) \bar{g}^{kl} D_l
A_{j_1 j_2 j_3}, 
\label{6:16}
\end{eqnarray}
with the covariant derivative $D_l A_{j_1 j_2 j_3}$, defined in 
Eq.(\ref{5:13}).
Putting
\begin{eqnarray}
dF_{\rm III} ^{(2)} &=& d \delta A^{(3)} = -*K^{(m-3)} , \nonumber \\
K^{(m-3)} &=& \frac{1}{(n-3)!} K_{i_1 i_2 ... i_{m-3}} du^{i_1} \wedge ...
\wedge du^{i_{m-3}} ,
\label{6:17}
\end{eqnarray}
one has the relation between the components of $F_{\rm III} ^{(2)}$ and
$K^{(m-3)}$:
\begin{equation}
F_{i_1 i_2 , i_3} + F_{i_2 i_3 , i_1} + F_{i_3 i_1 , i_2} = 
- \frac{1}{(m-3)!} \delta \left( \begin{array}{cccccc} 
1& 2& ...&...&...&m \\ j_1& ...& j_{m-3}& i_1& i_2& i_3 \end{array} \right)
\hspace{-1mm}\sqrt{\bar{g}} K^{j_1 ... j_{m-3}} ,
\label{6:18}
\end{equation}
where $F_{i_1 i_2 , i_3} \equiv \partial F_{i_1 i_2} / \partial\, u^{i_3}$, etc..
If our space-time $\bar{M}^m$ is {\it flat and the dimension is} $m=4$, these
expressions reduce to a familiar form.
\begin{eqnarray}
F_{\mu \nu} &=& -\partial ^{\rho} A_{\mu \nu \rho} , \nonumber \\
\tilde{F} _{\mu \nu} ,^{\nu} &=& K_{\mu} ,
\label{6:19}
\end{eqnarray}
where
\begin{eqnarray}
\tilde{F} _{\mu \nu} &=& \frac{1}{2} \epsilon _{\mu \nu \rho \sigma} F^{\rho \
sigma} , \nonumber \\
K &=& K_{\mu} du^{\mu} .
\label{6:20}
\end{eqnarray}
Equations (\ref{6:19}) and (\ref{6:20}) tell us that $K_{\mu}$
is a {\it magnetic monopole current}\cite{10}.

One finds, here also, that one stands on the viewpoint of tracing a monopole
by regarding the three form $A^{(3)}$ as a fundamental object.

The {\it gauge transformation} is, in this case, given by
\begin{eqnarray}
A^{(3)} \rightarrow \tilde{A}^{(3)} = A^{(3)} + \delta A^{(4)} .
\label{6:21}
\end{eqnarray}
In components is it written as
\begin{eqnarray}
\tilde{A}_{h_1 h_2 h_3} \hspace{-2mm}&=&\hspace{-2mm}A_{h_1 h_2 h_3}+
\frac{1}{4!(m-4)!}\delta\!
\left( \begin{array}{ccccccc}
h_1& h_2& h_3& l_1& ...&...& l_{m-3} \\ i_1& i_2& i_3& i_4& j_1& ...& j_{m-4} 
\end{array} \right) \nonumber \\
&&\times\frac{\partial}{\partial u^k} ( \sqrt{\bar{g}} A^{i_1 i_2 i_3 i_4})
\sqrt{\bar{g}}\hspace{1mm}
\bar{g}^{l_1 k} \bar{g}^{l_2 j_1} ... \bar{g}^{l_{m-3} j_{m-4}},
\label{6:22}
\end{eqnarray}
which one can further rewrite in the following form.
\[
\tilde{A}_{i_1 i_2 i_3}=A_{i_1 i_2 i_3} + \frac{1}{4!} \delta
\left( \begin{array}{cccc}
j_1 & j_2 & j_3 &j_4 \\ k & i_1 & i_2 & i_3 \end{array} \right)
g^{kl} D_l A_{j_1 j_2 j_3 j_4},
\]
\begin{equation}
D_l A_{j_1 j_2 j_3 j_4}=\frac{\partial A_{j_1 j_2 j_3 j_4}}{\partial u^l} -  A
_{k j_2 j_3 j_4} \Gamma _{j_1 l}^k - A_{j_1 k j_3 j_4} \Gamma _{j_2 l}^k - A_{
j_1 j_2 k j_4} \Gamma _{j_3 l}^k - A_{j_1 j_2 j_3 k} \Gamma _{j_4 l}^k .
\label{6:23}
\end{equation}

The action form ${\cal S}_{\rm III} = F_{\rm III}^{(2)} * F_{\rm III}^{(2)}$ 
can be, of course, calculated along the same line already mentioned.
And the total action ${\cal S}$ is
\begin{eqnarray}
{\cal S} = {\cal S}_{\rm I} + {\cal S}_{\rm II} + {\cal S}_{\rm III} .
\label{6:24}
\end{eqnarray}
\vspace{10mm}

\setcounter{equation}{0}
\section{Summary and conclusions}

\hspace*{6mm}
We have assigned a differential $q$-form $F^{(q)}$ to each point 
$x^{\mu}=x^{\mu} (u^i)$ of the submanifold $\bar{M}^m$ of the extended 
object's world, included in our observer's world $M^n$, thus endowing a 
particle {\it with an intrinsic degree of freedom}. 
An arbitrary $q$-form decomposes into a harmonic form, a $d$-boundary plus
a $\delta$-boundary. With $F^{(q)}$ we can make a scalar 
$(F^{(q)}, F^{(q)})$ defined by Eq.(\ref{2:2}) and we regard this as an action
for the system. 

Now, to say more concretely, if the assigned form is of zero, one is to 
have a generalized action for a point particle, a string or a $p$-brane in
a curved space-time.
The well-known Nambu-Goto action as well as the membrane action is 
thus naturally 
derived with this general principle.
Owing to the construction itself the action is reparametrization-invariant
and, at the same time, invariant under the general coordinate transformation.
For $q \geq 1$ we obtain a non-trivial action with spin degrees of freedom.
The case of $q=2$ is probably most attractive. The $d$-boundary 
$F_{\rm II} ^{(2)}$ has a fundamental 1-form $A^{(1)}$, and the world 
made of it is a conventional electromagnetic one, based 
on the Maxwell equation.
The $\delta$-boundary $F_{\rm III} ^{(2)}$ has, on the contrary, a 
fundamental 3-form $A^{(3)}$,
which is interpreted as a magnetic monopole current
(at least in case of $m=4$). 
Our equation differs from the conventional one only in that the former is
more general than the latter, if one admits the existence of monopoles.
The former is formulated on a general curved space-time with an arbitrary
space-time dimension.
In the conventional picture the space-time is flat.
So if one wants to construct the theory on curved space-time, one feels it
ambiguous to decide where to replace $\det(\delta _{ij})=1$ by 
$\det(g_{ij}) \neq 1$. 

Anyway one can construct an arbitrary $q$-form field over a general 
Riemannian manifold through de-Rham-Kodaira's theorem.
Thus , as usual, one is to assign spin degrees of freedom.

This last statement is important. A $p$-brane is usually considered as a
$p$-dimensional extended object $\bar{M}^{p+1}$ moving across our world
$M^n$. Each point of a $p$-brane has no internal degree of freedom.
On the contrary, if one takes up a $q$-form over $\bar{M}^m \subset M^n$,
one is to have an internal degree of freedom based on the number of 
components of the $q$-form.
The dimension of the local coordinate system $(u^1 , ..., u^m)$ of $\bar{M}^m$
indicates the dimension less by $1\,(p=m-1)$ of the extended object and the
dimension of the $q$-form represents the internal degree of each point.
A string is generated for $m=2$ and $q=0$, whereas a conventional 
$p$-brane, for $m=p+1$ and $q=0$.

Lastly we comment on the newly-introduced gauge transformation.
Gauge freedom comes from the nilpotency of the boundary operators 
$d$ and $\delta$ :
$d^2=0$ and $\delta ^2=0$, the latter of which induces a new type of
gauge transformations.
Equations(\ref{4:10}), (\ref{5:9}) and (\ref{6:21}) are such examples.
In the Dirac monopole theory with $n=m=4$ and $q=2$, we have a one-component
scalar $A^{(4)}$ which contributes to $A^{(3)}$, a monopole current.

Detail analysis along this way is worth studying and may promise a 
fruitful result about physical extended object.
\bigskip

The author thanks Prof. M.Wadati for researching facility at the University
of Tokyo.

\newpage
\addcontentsline{toc}{section}{Appendix}
\appendix
\renewcommand{\theequation}{\Alph{section}.\arabic{equation}}
\setcounter{equation}{0}
\section{Hodge's star operator}

\hspace*{6mm}
As defined by Eq.(\ref{2:1}), Hodge's star operator $*$ is an isomorphism of
${\cal H}^q$ (liner space of $q$-forms) into ${\cal H}^{m-q}$ .
Here, in this appendix, we only write down two important formulas which 
we frequently use in calculation in Sects.4 to 6.

For an arbitrary $q$-form
\begin{eqnarray}
\varphi = \frac{1}{q!} \varphi _{i_1 i_2 ... i_q} du^{i_1} \wedge du^{i_2}
\wedge ... \wedge du^{i_q}, 
\label{A:1}
\end{eqnarray}
we have
\begin{eqnarray}
* \varphi = \frac{1}{(m-q)!q!} \delta \left( \begin{array}{cccccc} 
1&2& ...&...&...& m \\ i_1 &... &i_q& j_1& ... &j_{m-q}  \end{array} \right) 
\sqrt{\bar{g}}\hspace{1mm}
\varphi ^{i_1 ... i_q} du^{j_1} \wedge ... \wedge du^{j_{m-q}}, 
\label{A:2}
\end{eqnarray}
where
\begin{eqnarray}
\varphi ^{i_1 ... i_q} = \bar{g} ^{i_1 l_1} ... \bar{g}^{i_q l_q} \varphi _{l_
1  ... l_q} ,
\label{A:3}
\end{eqnarray}
with $\bar{g}_{ij}$, the metric tensor. 

As for a basis of ${\cal H}^q$, we have 
\begin{eqnarray}
*(du^{k_1} \wedge ... \wedge du^{k_q}) = \frac{1}{(m-q)!} \delta \left( \begin
{array}{cccccc} 1 &2& ...&...&...& m \\ i_1&...& i_q& j_1& ...& j_{m-q}
\end{array} \right) \nonumber \\
\times \sqrt{\bar{g}}\hspace{1mm}\bar{g}^{i_1 k_1}...
\bar{g}^{i_q k_q} du^{j_1} \wedge ... \wedge du ^{j_{m-q}}. 
\label{A:4}
\end{eqnarray}
Note that a factor $1/q!$ is removed here in the right-hand side of Eq.(\ref{A:4}).


\begin{thebibliography}{99}
\bibitem{5}
	We will also call the local coordinate system by the name of the
	manifold itself.
\bibitem{6}
	Our manifold is assumed to be compact,
	so that harmonicity reduces to Eq.(2.4).
\bibitem{7}
	We are transforming a local coordinate system into another;
	remember the footnote \cite{5}.
\bibitem{8}
	Here, and henceforth, the components of the tensors
	$A_{{i_1}{i_2}...{i_n}}$ are always antisymmetric with
	respect to the exchange of suffices.
\end{thebibliography}
\end{document}